\documentclass[%
superscriptaddress,
preprint,
preprintnumbers,
amsmath,amssymb,pra,aps]{revtex4-2}

\usepackage[english]{babel}
\usepackage{amsmath}
\usepackage{graphicx}
\usepackage{float}
\usepackage{dcolumn}
\usepackage{bm}
\usepackage[colorlinks=true, allcolors=blue]{hyperref}
\usepackage[version=4]{mhchem}
\newcommand{\hmn}[1]{
  \ensuremath{\begingroup\setupHMN #1\endgroup}}
\newcommand{\setupHMN}{%
  \doHMN{-}{\HMNoverline}%
  \doHMN{*}{\HMNminverse}%
  \doHMN{i}{\infty}}
\newcommand{\doHMN}[2]{%
  \begingroup\lccode`~=`#1
  \lowercase{\endgroup\let~}#2%
  \mathcode`#1="8000}
\newcommand{\HMNminverse}[1]{\frac{#1}{m}}
\newcommand{\HMNoverline}[1]{\mkern1mu\overline{\mkern-1mu#1\mkern-1mu}\mkern1mu}
\begin{document}

\title{Toward machine learning interatomic potentials for modeling uranium mononitride}

\author{Lorena Alzate-Vargas}
\affiliation{Theoretical Division, Los Alamos National Laboratory, Los Alamos, New Mexico 87545, USA}%
\author{Kashi N. Subedi}
\affiliation{Theoretical Division, Los Alamos National Laboratory, Los Alamos, New Mexico 87545, USA}%
\author{Nicholas Lubbers}
\affiliation{Computer, Computational, and Statistical Sciences Division, Los Alamos National Laboratory}%
\author{Michael W.D Cooper}
\affiliation{Materials Science Division, Los Alamos National Laboratory, Los Alamos, New Mexico 87545, USA}%
\author{Roxanne M. Tutchton}
\affiliation{Theoretical Division, Los Alamos National Laboratory, Los Alamos, New Mexico 87545, USA}%
\author{Tammie Gibson}
\affiliation{Theoretical Division, Los Alamos National Laboratory, Los Alamos, New Mexico 87545, USA}%
\author{Richard A. Messerly}
\affiliation{Theoretical Division, Los Alamos National Laboratory, Los Alamos, New Mexico 87545, USA}%

\begin{abstract}
Uranium mononitride (UN) is a promising accident-tolerant fuel because of its high fissile density and high thermal conductivity. In this study, we developed the first machine learning interatomic potentials for reliable atomic-scale modeling of UN at finite temperatures. We constructed a training set using density functional theory (DFT) calculations that was enriched through an active learning procedure, and two neural network potentials were generated. Both potentials successfully reproduce key thermophysical properties of interest, such as temperature-dependent lattice parameter, specific heat capacity, and bulk modulus. We also evaluated the energy of stoichiometric defect reactions and defect migration barriers and found close agreement with DFT predictions, demonstrating that our potentials can be used for modeling defects in UN. Additional tests provide evidence that our potentials are reliable for simulating diffusion, noble gas impurities, and radiation damage. 
\end{abstract}

\maketitle

\section{Introduction}

The growing demand for safer nuclear energy has intensified research efforts toward accident-tolerant fuels. Among these, uranium mononitride (UN) has emerged as a particularly promising candidate due to its superior properties, including higher fissile density, enhanced thermal conductivity, and higher melting temperature compared to conventional UO$_2$ fuels. These advantages directly translate into improved reactor safety, enabling fuels to better withstand accidents involving coolant loss by reducing fuel center-line temperatures and delaying fuel melting events~\cite{nea2018}. 

Experimental evidence highlights several properties of UN, 
notably its compatibility with cladding materials and its relatively low rate of fission gas release compared to oxide fuels~\cite{Matthews1993, Watkins2021}. 
However, the adoption and licensing of a new accident-tolerant fuel requires a thorough understanding of the fuel, especially under realistic operating conditions to address the full range of material responses, which often requires extensive characterizations and crucial experimental measurements.

To address the challenge of limited experimental data, especially for novel fuels such as UN, multiscale multiphysics modeling frameworks have been increasingly utilized. These methods effectively reduce experimental workloads by leveraging computational tools to predict fuel performance over long operational durations. However, these models critically depend on accurate atomistic simulations, such as density functional theory (DFT) calculations and classical molecular dynamics (MD), to provide reliable estimates of material parameters.

Numerous DFT studies have extensively investigated the electronic structure and magnetic properties of uranium mononitride (UN)~\cite{Gryaznov2012, Mei2013, Kocevski2022a}. In addition, substantial efforts have been made to understand point defect behavior and the incorporation of fission products into the UN\cite{Kotomin2009, Ducher2011, Claisse2016, Szpunar2020, Yang2021, Zhao2023, Kuganathan2023}. These works have provided valuable insights into the mechanisms that govern fission gas migration and can further lead to fuel degradation.
Yet, despite their accuracy, DFT calculations are computationally demanding and typically limited to short timescales and small system sizes. In contrast, classical MD simulations enable exploration of larger systems and longer simulation times, for better prediction of fuel behavior at finite temperatures, but their reliability heavily depends on the quality of the interatomic potentials employed. 

For UN, several classical interatomic potentials have been developed with moderate success~\cite{Kuksin2016, Tseplyaev2016, Kocevski2022}.
Kuksin et al.~\cite{Kuksin2016} developed an angular-dependent potential, later refined by Tseplyaev and Starikov~\cite{Tseplyaev2016} to study defects, migration energies, and pressure-induced \hmn{fm-3m} $\rightarrow$ \hmn{R-3m} phase transitions of UN. Li and Murphy~\cite{Li2021} applied this potential to study self-diffusion in hypo-stoichiometric conditions. Another notable effort by Kocevski et al.~\cite{Kocevski2022} involved combining Buckingham-type interactions with the Finnis-Sinclair many-body embedded atom method potential, which can reproduce various thermal-dependent properties. As several differences arise between the available interatomic potentials, a robust and reliable UN potential is a necessity.

Recently, machine learning interatomic potentials (MLIPs) have been successfully employed to predict properties of nuclear fuels. For instance, Chen et al. applied a moment tensor potential to train a uranium MLIP capable of capturing multiple crystal phases~\cite{Chen2023}. Stippell et al. trained an MLIP for UO$_2$, showing the additional complexity introduced by DFT+\textit{U} calculations~\cite{Stippell2024} and moderately reproducing several temperature-dependent properties and reporting stoichiometric point defects. 
Dubois et al. demonstrated significant improvements in modeling the elastic and plastic behaviors of UO$_2$~\cite{Dubois2024}. While these studies demonstrate great potential and outline novel strategies for generating training datasets and efficiently sampling defect configurations, they also underscore significant challenges associated with modeling actinide materials. These challenges include the high computational expense inherent in generating sufficiently large datasets and difficulties in achieving consistent convergence within first-principles calculations, particularly when dealing with strongly correlated electronic systems like those involving actinide elements.

In this study, we have developed and validated two distinct MLIPs specifically tailored for UN: an \textit{accurate neural network engine for molecular energies} (ANI) neural network potential~\cite{Smith2017} and a \textit{hierarchically interacting particle neural network} (HIP-NN) potential~\cite{Lubbers2018, Chigaev2023}. Using a state-of-the-art active learning approach, our methodology ensures comprehensive sampling of relevant atomic configurations, including point defects. The resulting potentials exhibit excellent agreement with the DFT calculations in several thermophysical and thermomechanical properties, such as temperature-dependent heat capacity, defect formation energies, and defect migration barriers. Our results establish a reliable foundation for further exploration of atomic-scale phenomena in UN that can improve nuclear fuel performance models.

\section{Methods}

\subsection{Machine learning interatomic potentials}

We developed two MLIPs for uranium mononitride: an ANI-based neural network potential~\cite{Smith2017}  and a HIP-NN potential~\cite{Lubbers2018, Chigaev2023}. In a previous study, the ANI potential was used to develop an MLIP for UO$_2$ and since it was linked to the active learning framework used in this work it was our first choice.  However, recent studies have shown that HIP-NN can outperform ANI~\cite{Chigaev2023, Fedik2024}, therefore we decided to train a HIP-NN potential and examine the performance of both models across several material properties.

Similar to classical potentials, both MLIPs predict the total system energy as sums of atomic energies; atomic forces are obtained from gradients of the total energy with respect to atomic positions via automatic differentiation, ensuring energy conservation. 
Both ANI and HIP-NN  potentials incorporate only local atomic interactions within specified cutoff radii and guarantee invariance to translations, rotations, and atomic permutations, enabling transferable and physically consistent modeling of atomic interactions.\\ 

\textbf{ANI potential.}
ANI adopts the framework of Behler-Parinello neural network potentials~\cite{Behler2007}, utilizing radial and angular \textit{symmetry functions} to characterize the local atomic environment of each atom. Atomic environment vectors were constructed using both radial and angular symmetry functions. While the radial descriptors in ANI are consistent with the original Behler–Parrinello formulation, the angular symmetry functions are modified to improve the representation of local angular environments, thereby enhancing predictive accuracy for more complex atomic systems. 
In ANI, a separate neural network is trained for each atomic species, i.e., a separate neural network is trained for both uranium and nitrogen. The hyperparameters were adopted from prior optimization for UO$_2$~\cite{Stippell2024}, with the nitrogen-specific parameters adapted from those originally developed for oxygen. 

\textbf{HIP-NN potential.} 
HIP-NN utilizes an end-to-end, graph-based, message-passing neural network approach~\cite{Kocer2022}. Instead of predefined symmetry functions to describe the atomic environment, the characterization of the environment takes place through learned functions. In HIP-NN, this learning uses \textit{sensitivity functions} ($s$) which characterize distances between atoms, and \textit{interaction weights} ($V$) combining the distances and weights of neighboring atoms to form a message. The sensitivity functions are indexed by $\nu$ and take pairwise distance as input, the feature vector $Z$ is indexed by $a$ (output) and $b$ (input), and the neighbors of atom $i$ indexed by $j$. The crucial equation relating input features to the pre-activation of neurons, $A$, is
\begin{equation}
    \label{e1}
    A_{i,b} = \sum_{\nu,j,b} V^\nu_{ab} s^\nu(r_{ij}) Z_{j,b}\;.
\end{equation}

The relation in Eq.~(\ref{e1}) demonstrates that the message received is a learned-product reduction between the sensitivity functions and the features of neighbors, summed over all neighbors in the environment of $i$. In contrast to the species-based symmetry functions of Behler-Parinello networks, this equation can be applied recursively using multiple interaction layers to generate deeper and longer-range interactions.
We used the \emph{hippynn}~\cite{hippynn} open-source implementation of HIP-NN for this work, which provides not only functions for learning, but also an interface to molecular dynamics simulations with Large-scale Atomic/Molecular Massively Parallel Simulator (LAMMPS) package~\cite{Thompson2022}.\\

The ANI and HIP-NN model hyperparameters can be found in Appendix~\ref{app:hyper}.\\

\subsection{Active learning procedure for dataset generation}

\begin{figure*}[ht]
\includegraphics[width=1\linewidth]{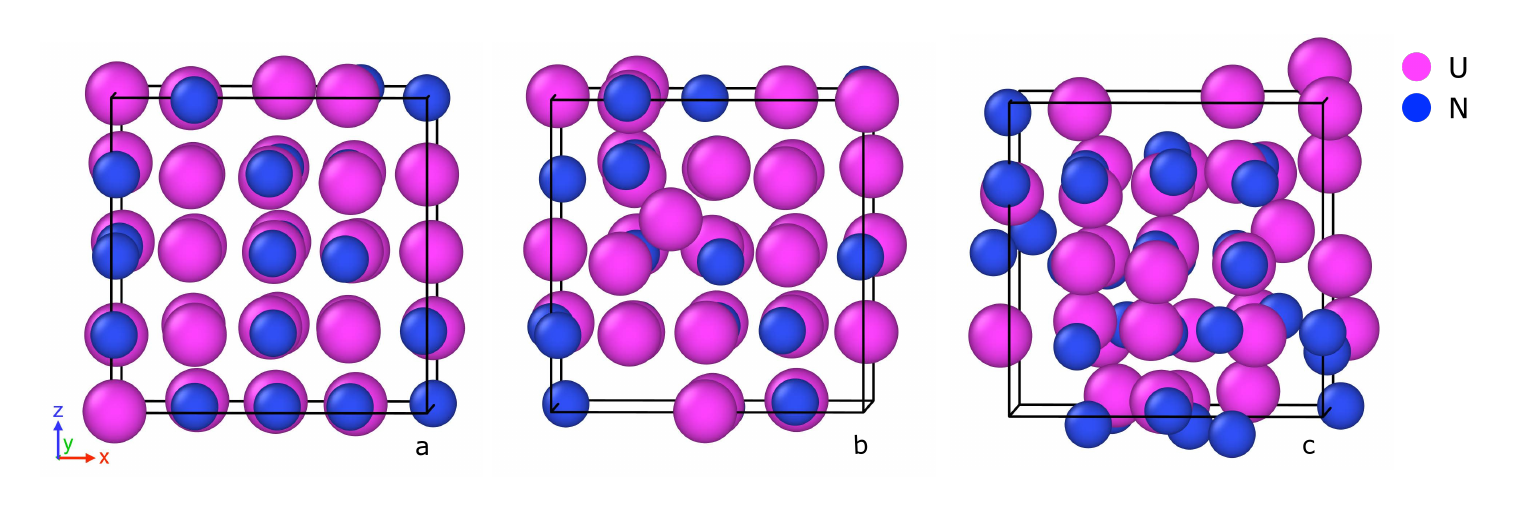} 
\caption{\label{fig:snaps} Snapshots of UN atomic configurations found in the training dataset consisting of (a) stable crystalline rock salt structure from the bootstrap dataset, (b) highly disordered structure from the last active learning iteration dataset, and (c) structure containing an uranium interstitial that is part of the last active learning iteration dataset.}
\end{figure*}

The accuracy of MLIPs depends critically on the quality, diversity, and coverage of their training datasets. To construct such datasets efficiently, active learning strategies have been developed to autonomously explore chemical space and iteratively improve model performance. These approaches identify and incorporate atomic configurations where model predictions are uncertain or inaccurate, thus increasing the robustness of the model. New configurations are generated via low-cost \textit{sampling} methods --typically MD simulations-- and a selected subset are then \textit{labeled} using first-principles calculations to obtain reference energies and forces that will be added to the ever-growing dataset.
A widely adopted selection strategy is query-by-committee (QBC), in which an ensemble of independently trained MLIP models is used to estimate uncertainty~\cite{Smith2018}. For each candidate configuration, disagreement among the ensemble’s predictions is quantified on both the total energy and atomic forces.

Initially, we created a bootstrap dataset containing configurations of 64-atom supercells, primarily in stable crystalline rock salt (\hmn{fm-3m}) structures as shown in Fig.~\ref{fig:snaps}(a), along with configurations involving small perturbations of the stable structure. Initial labeling was performed by selecting random configurations from this set and calculating their DFT energies and forces from static calculations. Additionally, to enhance diversity, configurations representing defect types--neutral Schottky defects, uranium and nitrogen Frenkel pairs--were manually added to the dataset in the early iterations of AL and subsequently used for sampling. Fig.~\ref{fig:snaps}(b) shows an atomic snapshot of a structure with an uranium at an interstitial position, sampled in the last iteration of AL and included in the dataset.

Atomic configurations were sampled through MD simulations employing the MLIP ensemble for the current AL iteration, applying systematic temperature and density perturbations to explore a broad range of configurations, including non-equilibrium states relevant to high-temperature or high-pressure environments~\cite{Smith2018}. Structures become increasingly disordered with AL iterations as can be seen from the snapshot in Fig.~\ref{fig:snaps}(c) of a highly disordered configuration obtained in the last iteration. Full details of the sampling parameters, including temperature and density ranges are provided in Appendix~\ref{app:al}.
 
Only high-uncertainty configurations from the MD sampling simulations are included in the training dataset during AL. To estimate the model uncertainty, an ensemble of eight ANI-type MLIPs was trained using randomized data splits (training, validation, test) and different initial model parameters. We employed ANI for AL because of the higher speeds for both learning in training and inference in sampling.

The final training dataset used to train both ANI and HIP-NN potentials contains 12,336 atomic configurations, covering crystalline, defect-containing, and disordered structures.

\subsection{Density functional theory calculations}

All DFT calculations were performed using plane-wave-based Vienna \textit{ab initio} simulation package (VASP 6.3.2). The generalized-gradient approximation (GGA)~\cite{Perdew1996} with projected-augmented wave (PAW)~\cite{Kresse1999} potentials was used to describe the exchange-correlation interactions. Spin-polarized calculations with ferromagnetic (FM) ordering on uranium atoms were performed as FM is energetically favorable with the PBE functional. Since the MLIPs will be applied to predict physical properties that are highly sensitive to lattice dynamics, we benchmarked the effect of exchange-correlation functionals by calculating the phonon dispersion of UN. Specifically, we compared the PBE and PBE+$U$~\cite{Dudarev1998} 
functionals (see Appendix~\ref{app:dft}). Among both, PBE was the only functional that did not yield imaginary (i.e., negative) phonon frequencies, indicating dynamic stability of the crystal structure. Consequently, PBE was selected as the most reliable and physically consistent choice for generating reference data to label our UN training set.

Labeling calculations (energy and forces) used a $2 \times 2 \times 2$ supercell (64 atoms or 62 atoms for Schottky defect configurations) of the rock salt unit cell with periodic boundary conditions. We set a cutoff energy of 520 eV to expand the Kohn-Sham orbitals into the plane-wave basis sets, and the Methfessel-Paxton smearing method with width of 0.1 eV was applied~\cite{Methfessel1989}.
Convergence criteria were set to $10^{-8}$ eV for energy and $10^{-4}$ eV/\AA~for atomic forces, with the Brillouin zone sampled using a $3 \times 3 \times 3$ \textit{k}-point grid.

\subsection{Molecular dynamics simulations}

Molecular dynamics simulations for active learning were performed using the Atomic Simulation Environment (ASE)~\cite{ASE}. All other MD simulations were performed using LAMMPS~\cite{Thompson2022}. Periodic boundary conditions were applied in all cases. 

Elastic constants and defect energetics at 0 K were computed using a $5 \times 5 \times 5$ supercell via finite deformation methods and nudged elastic band (NEB) calculations for migration barriers.
Phonon spectra were calculated using the finite displacement method implemented in ASE using a $10 \times 10 \times 10$ supercell, and atomic displacements of 0.05 \AA~to compute the dynamical matrix.

Finite-temperature properties were evaluated using a $15 \times 15 \times 15$ supercell under the isothermal-isobaric (NPT) ensemble, employing Nos\'e–Hoover thermostats and barostats with damping parameters set at 0.1 ps and 0.5 ps, respectively. 
Simulations spanned temperatures from 300~K to 2700~K at intervals of 25~K, each lasting 30 ps with a timestep of 1 fs. Properties such as lattice parameters, specific heat capacities--calculated from the temperature derivative of enthalpy-- and bulk moduli --determined from the derivative of pressure with respect to volume under compressive and tensile strains-- were derived from equilibrium averages over the final 10 ps of each run. 

Nitrogen self-diffusion was studied using the HIP-NN potential by analyzing the mean square displacements data from a 1-ns simulation under microcanonical (NVE) conditions at moderate to high temperatures (1800~K to 2800 K) followed initial equilibration under NPT conditions. Hypo-stoichiometric scenarios with controlled nitrogen vacancies and interstitials were also simulated to investigate defect-assisted diffusion mechanisms.

Xenon impurity simulations were performed using the HIP-NN potential. Because our training dataset does not contain Xenon, we used the Buckingham potential parameters developed by Kocevski et al.~\cite{Kocevski2022} to describe Xe–U and Xe–N interactions.

The ANI potential internally implements an ensemble-averaging approach when interfaced with LAMMPS, inherently providing averaged results from a single MD run. Conversely, the HIP-NN potential yields multiple distinct model instances. Thus, for properties calculated with HIP-NN, we conducted eight independent simulations, from which average values and standard deviations were computed.

Collision cascade simulations were performed using the HIP-NN potential for a large system containing approximately four million atoms ($80 \times 80 \times 80$ unit cells). Cascades were initiated by imparting a kinetic energy of 1 keV to a randomly oriented uranium primary knock-on atom (PKA). Initially equilibrated under NPT conditions at 600 K for 20 ps and zero pressure, the cascade simulation subsequently proceeded under NVE conditions with adaptive timesteps capped at 1 fs. Defect analysis--to identify interstitials, vacancies, and non-defective lattice sites--was done using the Wigner–Seitz cell method in the Open Visualization Tool (OVITO) software~\cite{ovito}.

\section{Results}

\subsection{Validation of models}

To evaluate the performance of the ANI and HIP-NN potentials developed for UN, we assess their accuracy using the root mean square error (RMSE) for total energies and atomic forces across the entire dataset (Table~\ref{tab:rmse}). The final training dataset consists of $2 \times 2 \times 2$ supercells containing approximately 64 atoms, with slight variations due to defect formation. Both potentials demonstrate robust predictive capability, given the dataset complexity that encompasses various temperature regimes and defect structures, 
and compare favorably with existing MLIPs in the literature. For example, the energy and force RMSEs are quite similar to those reported for the  UO$_2$ ANI potential trained to DFT (2.17 meV/atom and 0.091 eV/\AA).\\

\begin{table}[ht!] 
\caption{\label{tab:rmse}Training root mean square error (RMSE) values for energy (meV/atom) and atomic forces (eV/\AA) calculated with ANI and HIP-NN potentials.}
\begin{ruledtabular}
\begin{tabular}{ccc}
 \textbf{RMSE}    & \textbf{ANI}  & \textbf{HIP-NN}  \\
   \hline 
Energy (meV/atom) & 2.33   &  2.48 \\
Force (eV/\AA)    & 0.072  &  0.113 \\
\end{tabular}
\end{ruledtabular}
\end{table}

In addition, we compare the lattice parameters at 300~K ($a_{300}$), elastic constants, and bulk moduli at 0~K predicted by the MLIPs to experimental data, DFT (PBE), and the classical potentials by Tseplayaev and Kocevski (Table~\ref{tab:elastict}). Our MLIPs demonstrate very good agreement with the DFT reference values calculated in this work, indicating that both ANI and HIP-NN potentials effectively capture the essential physics provided by the training data. When compared to the experimental data, we find that the MLIPs predict $a_{300}$ (error $\sim$0.7\%) and bulk modulus (error $\sim$2.4–7.3\%) accurately. However, larger discrepancies emerge for certain elastic constants, particularly $C_{44}$ (error $\sim$39–43\%), that partially stem from limitations inherent to the PBE exchange-correlation functional. Despite not being trained directly to experimental data, our MLIPs demonstrate comparable accuracy to classical potentials by Tseplyaev and Kocevski, which were fitted to reproduce DFT calculations and, in the case of Kocevski, also experimental properties.
\begin{table*}[ht!] 
\caption{\label{tab:elastict}Lattice parameter at 300 K ($a_{300}$), elastic constants ($C_{ij}$) and bulk modulus ($B$) at 0~K calculated using the ANI and HIP-NN potentials.  Experimental data, DFT calculations from this work using PBE functional and estimated values using classical potentials by Tseplayaev and Kocevski are included for comparison.}

\begin{ruledtabular}
\begin{tabular}{cccccc}
         & \textbf{$a_{300}$ (\AA)}  & \textbf{$C_{11}$ (GPa)}  & \textbf{$C_{12}$ (GPa)}  & \textbf{$C_{44}$ (GPa)}  & \textbf{$B$  (GPa)} \\
   \hline 
ANI     		       & 4.857              & 393      & 119      & 46       & 211 \\
HIP-NN   			   & 4.858              & 399      & 131      & 43       & 221 \\
Exp~\cite{Salleh1986}  & 4.89               & 423      & 98       & 75       & 206 \\
DFT (this work)   & 4.86\footnote{0 K} & 405      & 127      & 52       & 219 \\
Tseplyaev			   & 4.809              & 586      & 110      & 55       & 269 \\
Kocevski			   & 4.897              & 425      & 117      & 71       & 220
\end{tabular}
\end{ruledtabular}
\end{table*}

\begin{figure}[ht] 
    \centering
    \includegraphics[width=0.65\linewidth]{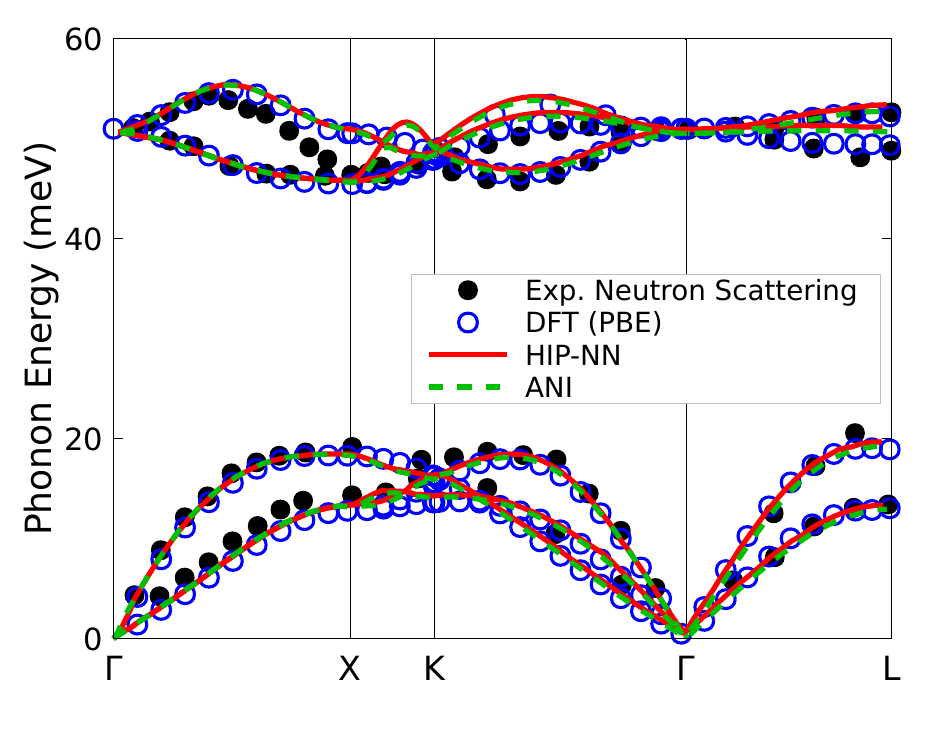}
    \caption{\label{fig:phonon}UN phonon band dispersion calculated by MLIPs. Experimental data~\cite{Jackman1986} and DFT (PBE) data have been added for comparison.}
\end{figure}

Phonon band structures calculated with both MLIPs are shown in Fig.~\ref{fig:phonon}. For comparison, experimental data (black empty points) and DFT spectra calculated with the same PBE reference method (blue empty points) are included. The agreement is excellent between our MLIPs and the DFT spectra. When comparing to experimental measurements at 4.2 K made by Jackman et al.~\cite{Jackman1986}, both ANI and HIP-NN accurately reproduced acoustic modes (lower energies), which is generally achievable even by classical potentials--see phonon spectra calculated with classical potentials in Ref.~\cite{AbdulHameed2024}). However, both MLIPs significantly improve the predictions of optical modes (higher energies), reflecting superior accuracy in modeling vibrations compared to classical potentials.\\

Since the incorporation of impurities in UN is facilitated by point defects, we validate the accuracy of the defect formation energies of uranium Frenkel pair (UFP), nitrogen Frenkel pair (NFP), and bound/unbound Schottky defects (SD). As summarized in Table~\ref{tab:defect}, our MLIPs provide excellent predictions, closely matching the DFT reference calculations obtained in this work and surpassing the classical potentials previously developed~\cite{Tseplyaev2016,Kocevski2022}, particularly for uranium Frenkel pairs. Previous studies show that the Kocevski potential does not accurately predict the formation energies for U-rich and stoichiometric conditions due to the inability to describe metallic uranium~\cite{AbdulHameed2024}.

\begin{table}[ht!]
\caption{\label{tab:defect}Formation energies (eV) for stoichiometric point defects obtained from MLIPs and compared with DFT (PBE) and classical potentials by Tseplyaev and Kocevski.}
\begin{ruledtabular}
\begin{tabular}{ccccc}
      & \textbf{UFP} & \textbf{NFP} & \textbf{SD$_\text{bound}$}  & \textbf{SD$_\text{unbound}$} \\
   \hline
ANI           & 9.87  & 5.01 & 4.49 & 4.86 \\
HIP-NN        & 9.86  & 4.99 & 4.7  & 5.16 \\
DFT (this work) & 9.46  & 5.04 & 4.47 & 5.15 \\
Tseplyaev     & 9.32  & 4.46 & 4.33 & 4.52 \\
Kocevski      & 14.37 & 4.00 & 2.99 & 3.89 \\
\end{tabular}
\end{ruledtabular}
\end{table}

\subsection{Thermodynamic properties}

Having verified the reliability of each MLIP for 0~K lattice properties and point defects, we now investigate the performance on thermodynamical properties at finite temperatures.

\begin{figure}[ht!] 
    \centering
    \includegraphics[width=0.65\linewidth]{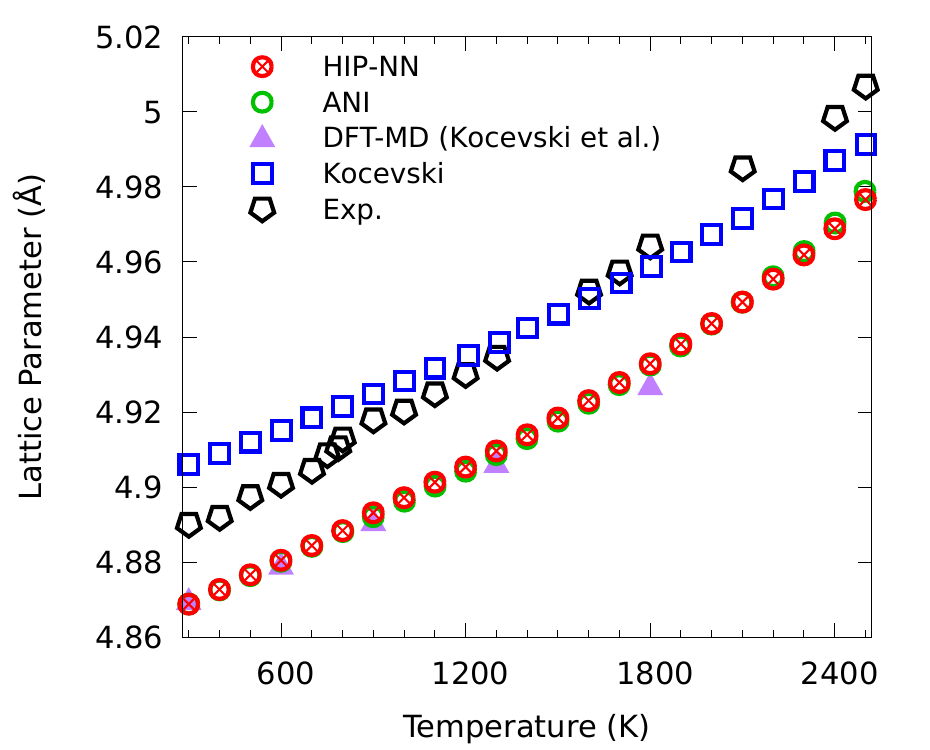}
    \caption{\label{fig:lattice}Lattice parameter as a function of temperature for both MLIPs, in comparison with experimental data from Hayes et al.~\cite{Hayes1990_1}, DFT-MD data~\cite{Kocevski2023} and classical potential prediction from Koceveski et al.~\cite{Kocevski2022}}
\end{figure}

The lattice parameter as a function of temperature is shown in Fig.~\ref{fig:lattice}. We compare the MLIP-predicted lattice parameter with experimental data, finite-temperature DFT-MD data calculated for an ferromagnetic UN structure using the PBE functional~\cite{Kocevski2023}, and classical potential MD simulation results. Our MLIPs systematically underestimate (by less than 1\%) the experimental values across the temperature range, yet capture the correct thermal expansion trend. The excellent agreement of both MLIPs with respect to the DFT-MD data indicates that the offset with respect to experiment comes mainly from the chosen PBE functional rather than the MLIP architecture, inadequate data coverage, or poor training.

\begin{figure}[ht!]
    \centering
    \includegraphics[width=0.65\linewidth]{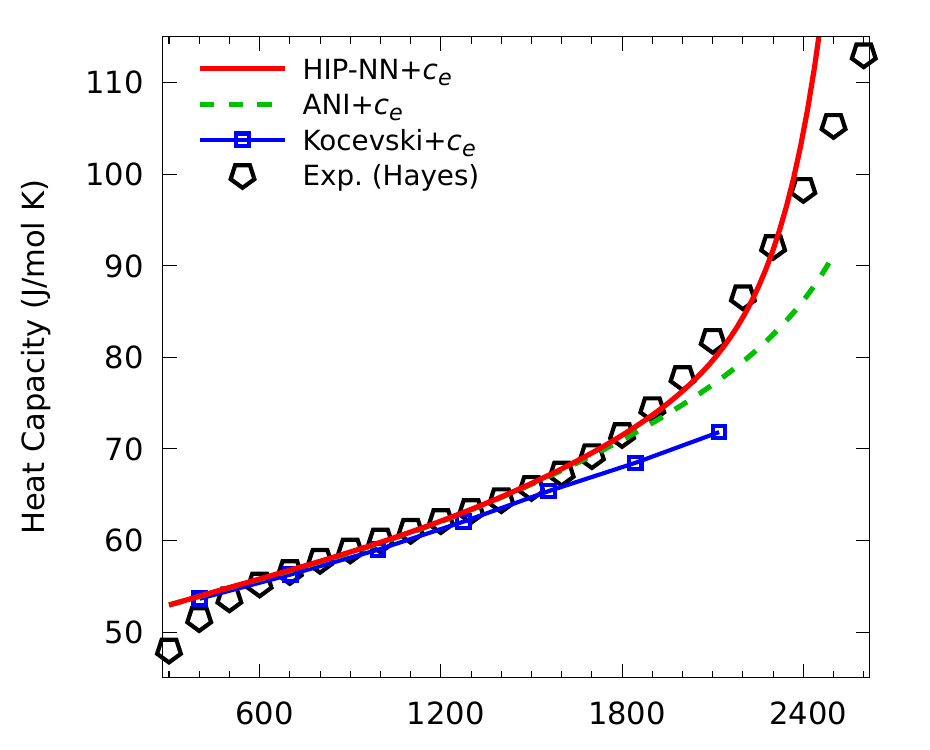}
    \caption{\label{fig:heat}Specific heat capacity as a function of temperature for both MLIPs, in comparison with experimental data (open symbols) from Hayes et al.~\cite{Hayes1990_4} and the classical potential prediction by Kocevski et al.~\cite{Kocevski2022}. The electronic contribution for ferromagnetic UN ($c_e$), calculated using DFT from Ref.~\cite{Szpunar2020} has been included in all MD-derived heat capacities.}
\end{figure}

The specific heat capacity as a function of temperature is shown in Fig.~\ref{fig:heat}. Since MD calculations only capture the vibrational component of the heat capacity—excluding electronic and magnetic contributions--we included electronic corrections ($c_e$) derived from DFT calculations (Ref.~\cite{Szpunar2020}). With these corrections, both MLIP-based predictions were compared to experimental data from Hayes et al.~\cite{Hayes1990_4} and to the classical potential from Kocevski et al.
At lower temperatures, where vibrational contributions dominate, all models--including the classical potential--show good agreement with experimental data. This consistency confirms that the vibrational baseline is well-represented and that the inclusion of DFT-based electronic corrections is essential for matching absolute values.
 
At higher temperatures, experimental data reveal an enhanced heat capacity, attributed to thermally activated point defects. This behavior, often modeled via an exponential defect term, has been observed in several actinide materials, including UO$_2$ and UC, where defect generation becomes increasingly significant at elevated temperatures~\cite{Pavlov2017}.
In this high-temperature regime (above 1500~K), both MLIPs outperform the classical Kocevski potential. The HIP-NN model, in particular, shows excellent agreement with experimental data, successfully capturing the sharp rise in specific heat associated with defect contributions. The ANI potential also reproduces the increasing trend in heat capacity, though its agreement with experiment is somewhat less accurate than that of HIP-NN. Given that both models were trained on the same datasets, this discrepancy likely arises from differences in model architecture or hyperparameter optimization.

\begin{figure}[ht] 
    \centering 
    \includegraphics[width=0.65\linewidth]{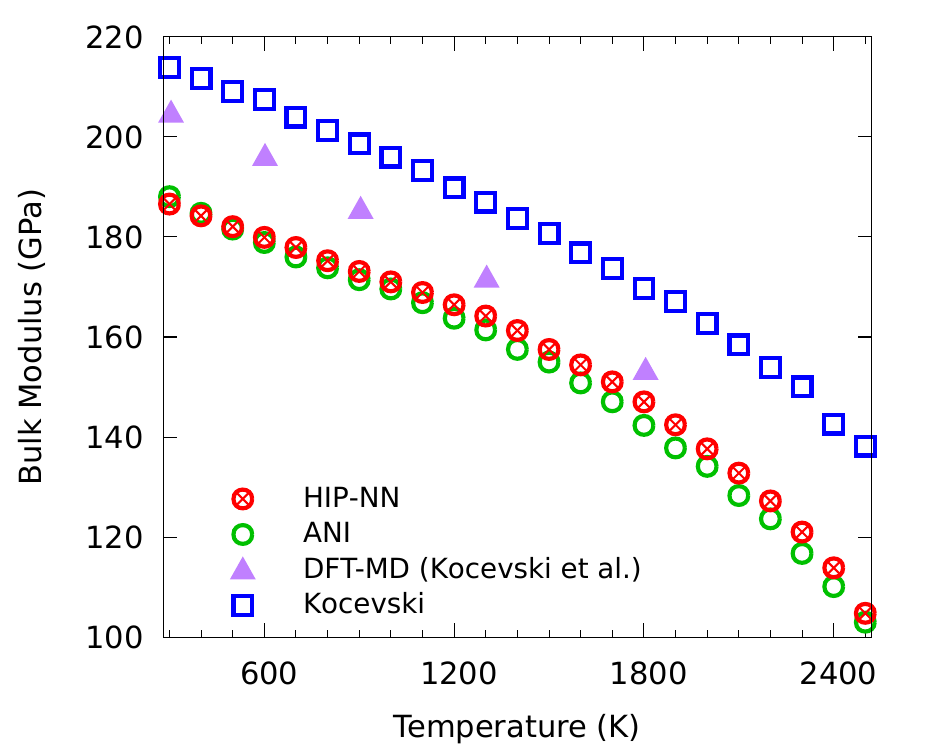}
    \caption{\label{fig:bulk}Bulk modulus as a function of temperature for both MLIPs, in comparison with DFT-MD data~\cite{Kocevski2023} and classical potential prediction by Kocevski et al.~\cite{Kocevski2022}}
\end{figure}
The bulk modulus as a function of temperature is shown in Fig.~\ref{fig:bulk}. Since experimental data are not available for comparison -- bulk modulus data by Hayes et al.~\cite{Hayes1990_2} were inferred from the UN porosity --we compare our estimated bulk modulus with DFT-MD using PBE functional and classical potential predictions. Our MLIPs correctly predict that the bulk modulus exhibits an inverse relation with lattice expansion. Although our MLIPs slightly underestimate bulk modulus values at lower temperatures relative to the reference data, high-temperature predictions aligned closely with existing computational references. 

\subsection{Defect migration barriers, diffusion, and Xe incorporation energies}

Accurate predictions of self-diffusion and impurity incorporation energies, such as Xenon, are crucial for understanding nuclear fuel behavior. These properties help elucidate the irradiation behavior of nuclear fuels. As the results for ANI and HIP-NN are quite similar, we focus this section solely on the HIP-NN model.

NEB migration barriers for nitrogen and uranium defects: vacancies (vac) and interstitials (int) are given in Table~\ref{tab:neb}.
Although our sampling scheme/dataset generation approach does not directly target migration barriers, the high-temperature simulations with interstitial defects in the initial structure are intended to explore these rare events. HIP-NN is able to provide reliable predictions that are in moderate agreement with the DFT reference data (calculated using PBE on a $2 \times 2 \times 2$ supercell). While improvement would require direct NEB sampling along migration pathways, the MLIP predicts a better migration barrier for U$_\text{int}$ than the classical potential of Kocevski et al.~\cite{Kocevski2022}\\

\begin{table} [ht!]
\caption{\label{tab:neb} Migration barriers (eV) for nitrogen (N) and uranium (U) vacancies (vac) and interstitials (int). For comparison, DFT (PBE) calculations~\cite{Kocevski2022} and classical MD predictions are included.}
\begin{ruledtabular}
\begin{tabular}{ccccc}
    & \textbf{N$_\text{vac}$} & \textbf{U$_\text{vac}$}  & \textbf{N$_\text{int}$}  & \textbf{U$_\text{int}$} \\
   \hline 
HIP-NN          & 2.689 & 2.894 & 1.69  & 0.66 \\
DFT (Ref.~\cite{Kocevski2022})  & 2.983 & 3.396 & 2.422 & 0.976 \\
Kocevski        & 2.524 & 3.363 & 2.012 & 0.28 \\
\end{tabular}
\end{ruledtabular}
\end{table}

\begin{figure}[ht!] 
   \centering
    \includegraphics[width=0.65\linewidth]{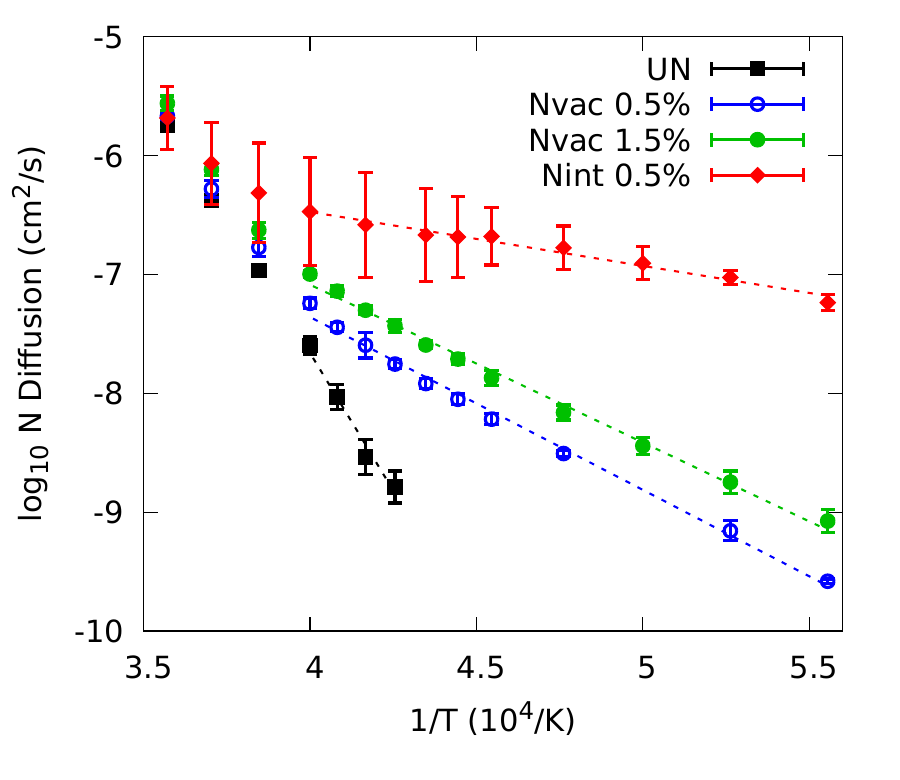}
    \caption{\label{fig:dif} Nitrogen self-diffusion vs inverse temperature using HIP-NN for an stoichiometric UN, a system with 0.5\% nitrogen vacancies and 1\% nitrogen vacancies, and a UN system with 0.5\% nitrogen interstitials. Error bars are standard deviation from eight simulations with distinct ensemble members.}
\end{figure}

We investigate nitrogen self-diffusion in both stoichiometric and hypo-stoichiometric UN using the HIP-NN potential. Diffusion coefficients computed across a wide temperature range are summarized in Fig.~\ref{fig:dif}. In stoichiometric UN, measurable diffusion only emerges above 2300~K, consistent with the expected high migration barrier for nitrogen in a defect-free lattice.

To explore the role of point defects, we study two hypo-stoichiometric compositions with nitrogen vacancy concentrations of 0.5\% and 1\%, as well as a hyper-stoichiometric case with 0.5\% nitrogen interstitials. Compared to pure UN, both vacancy concentrations yielded systematically higher diffusivities and significantly lower activation energies. 
The activation energy for nitrogen diffusion in hypo-stoichiometric UN at low temperatures was found to be 2.82~eV for 0.5\% vacancies and 2.57~eV for 1\% vacancies. As expected, increasing the vacancy concentration lowers the barrier due to a vacancy-assisted hopping mechanism. 

The hyper-stoichiometric interstitial system exhibited even greater enhancement. The much lower activation energy of 0.97~eV indicates a fundamentally different and more efficient diffusion mechanism. At higher temperatures, there is an increased statistical uncertainty between MLIP ensemble members, likely reflecting the larger configurational space accessible to nitrogen interstitials and their potential to undergo correlated or burst-like migration events. 

The diffusion results demonstrate that nitrogen transport in UN is strongly facilitated by both vacancy and interstitial defects, with interstitials emerging as the dominant diffusion pathway under non-stoichiometric conditions. This trend aligns with experimental observations suggesting that self-diffusion in uranium mononitride proceeds primarily via interstitial-mediated mechanisms\cite{Holt1969, Matzke1990}.

At high temperatures (above 2500~K), the diffusion coefficients of all systems--stoichiometric and non-stoichiometric--begin to converge. This convergence suggests that thermal energy becomes sufficient to overcome even high migration barriers, activating diffusion pathways regardless of initial defect concentration.

\begin{table}[ht!]
\caption{\label{tab:xe} Incorporation energies of Xenon (eV) in UN in comparison to reference DFT and classical potential data provided by~\cite{Kocevski2022}.}
\begin{ruledtabular}
\begin{tabular}{cccc}
    & \textbf{Xe$_\text{int}$} & \textbf{Xe$_\text{N}$} & \textbf{Xe$_\text{U}$} \\
   \hline 
HIP-NN          & 16.85 & 8.96 & 4.0  \\
DFT (Ref.~\cite{Kocevski2022}) & 14.68 & 8.45 & 4.84 \\
Kocevski        & 16.39 & 7.89 & 3.96 \\
\end{tabular}
\end{ruledtabular}
\end{table}

Xenon is a critically important fission product for the performance of nuclear fuel, as it precipitates to form bubbles which can induce fuel swelling. While studying fuel swelling was not the focus of this study, we report the Xe incorporation energies for different sites in Table~\ref{tab:xe}.
Our predicted incorporation energies of Xe into a nitrogen vacancy (Xe$_\text{N}$) and a uranium vacancy (Xe$_\text{U}$) are in excellent agreement with reference DFT (PBE) data calculated by Kocevski et al.~\cite{Kocevski2023} However, the incorporation energies for a Xe atom at the tetrahedral interstitial site (Xe$_\text{int}$) are overestimated by more than 2 eV with our hybrid HIP-NN and Buckingham potential approach. While the accuracy of the incorporation energies are likely limited in accuracy by the Buckingham potential used for Xe, these results demonstrate that our MLIP is compatible with simple classical potentials for simulating noble gas impurities.

\subsection{Cascade reactions}

Cascade simulations are widely used to investigate the atomic-scale effects of irradiation in materials. Although detailed irradiation damage studies were not the primary objective of this work, we demonstrate that our HIP-NN potential is capable of performing large-scale cascade simulations for modeling defect formation processes in UN.

\begin{figure*}[ht!]
\includegraphics[width=1\linewidth]{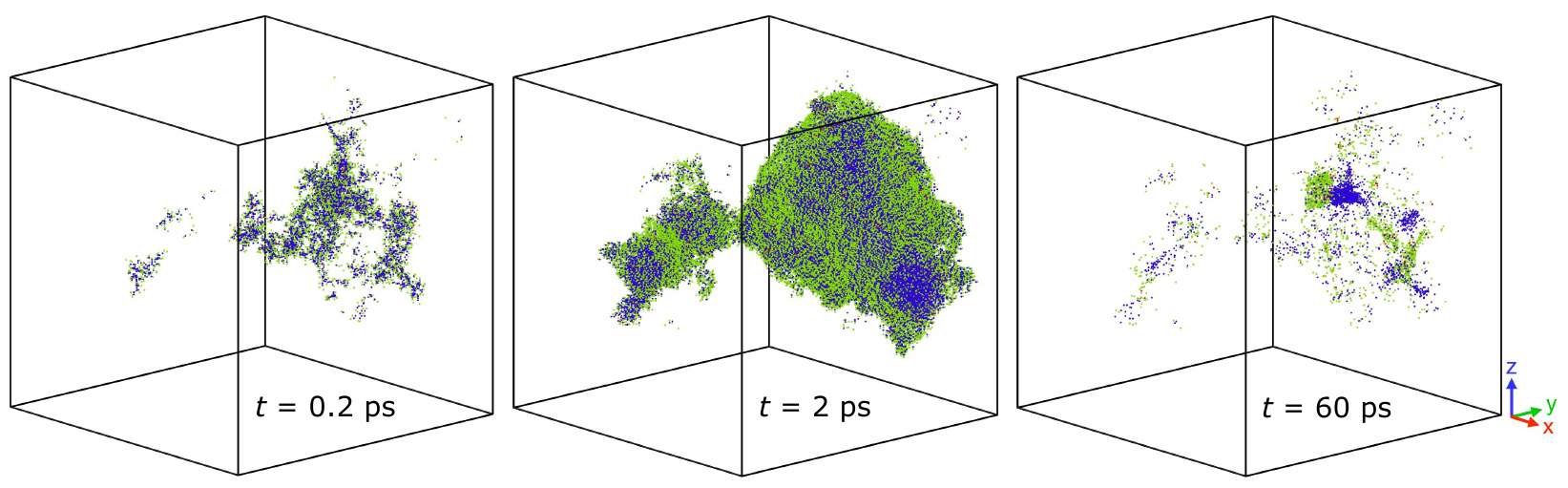}%
\caption{\label{fig:cascade} Snapshots of sites containing interstitial defects or vacancies during a collision cascade.}
\end{figure*}

We simulate a 1 keV collision cascade initiated by a uranium primary knock-on atom (PKA). The impact triggers a sequence of atomic displacements and local disorder. We monitor the evolution of point defects over time, focusing on the formation of Frenkel pairs--composed of interstitial-vacancy pairs--as a key indicator of damage.
Figure~\ref{fig:cascade} shows representative snapshots of the cascade evolution. At 0.2 picoseconds after impact, significant Frenkel pair formation occurs near the PKA origin. The number of defects peaks around 2 picoseconds and then gradually declines as atoms recombine or return to lattice positions. By several tens of picoseconds, the system reaches a steady state, with the surviving defect population stabilized.

While this single simulation is not intended as an in-depth study of radiation effects, it provides evidence that our MLIP can serve a broader applicability for modeling radiation damage.

\section{Conclusions}

We present two machine learning interatomic potentials that can describe uranium mononitride for studying atomic-scale phenomenon and for predicting physical properties. As shown in previous studies, developing a machine learning interatomic potential for nuclear fuels can be a challenging task, especially for actinide compounds. 

First, each potential was validated with the 300~K lattice parameter, 0~K elastic constants, 0~K bulk modulus and phonon spectra. Further validation investigated the temperature-dependent lattice parameter, specific heat capacity and bulk modulus over a temperature range from 300~K to 2700~K. Both potentials are capable of accurately reproducing DFT-MD simulation results and experimental data, when available.


The potentials were also validated by evaluating defect energies, key for a good understanding of the nuclear fuel behavior under irradiation. We find that both developed potentials demonstrate excellent agreement with reference DFT calculations for point defect energies. 

The HIP-NN trained potential was also used to calculate migration barriers for several defects. Despite these rare events not being included in the training dataset, HIP-NN predictions show reasonable agreement with DFT calculations. HIP-NN was also used to study diffusion coefficients and diffusion mechanisms, achieving results consistent with experiment. Xenon incorporation energies, calculated with a hybrid HIP-NN and Buckingham potential model, are also in close agreement with DFT. Finally, we demonstrate the capability of HIP-NN to perform large-scale cascade simulations. 

Future work will explore additional properties, such as vibrational entropy, to further enhance multiscale simulation capabilities for fuel performance. We also anticipate applying other techniques, such as experimental refinement to improve the agreement with experiment.

\begin{acknowledgments}
The authors appreciate the helpful scientific discussions with Benjamin Nebgen, Justin Smith, Shriya Gumber, Sakib Matin, and Anton Schneider. Los Alamos National Laboratory (LANL), an affirmative action/equal opportunity employer, is operated by Triad National Security LLC, for the National Nuclear Security Administration of the U.S. Department of Energy under Contract No. 89233218CNA000001. Research presented in this work was supported by the Laboratory Directed Research and Development (LDRD) program via project 20220053DR at LANL. We also appreciate the resources provided by the LANL Institutional Computing (IC) program.
\end{acknowledgments}

\section*{Data Availability}
The active learned training dataset and final ANI and HIP-NN potentials will be available after completing LANL reviewing process.

\section*{Appendix}
\appendix

\section{MLIP Hyperparameters}\label{app:hyper}

\textbf{ANI potential.} The neural networks for U and N are composed of three hidden layers with 96, 96, and 64 nodes, using the CELU activation function and a linear output layer. The radial component employed a cutoff of 7.0 \AA, with 32 Gaussian radial basis functions centered between 1.25 \AA~and 6.82 \AA. The angular descriptor used a radial cutoff of 5.0 \AA, with 64 basis functions composed of 8 radial shifts and 8 angular shifts, distributed across interatomic distances and angles. ANI ensemble was trained using both energy and force labels, with respective loss weights of 1.0 and 0.1.

\textbf{HIP-NN potential.} In this work, we found that a single interaction layer was sufficient ($n_\textrm{interaction}=1$). 
We used $n_\textrm{feature}=60$ neurons per layer and $n_\nu=20$ sensitivity functions.
We used the improved interaction HIP-NN form that includes tensor sensitivities~\cite{Chigaev2023} by setting the tensor order ${\ell}=1$, that is, vector-based message passing, which improves over the original HIP-NN by summing messages generated from neighbors not only as scalars but also as vectors. This allows individual neurons to be sensitive to angles between neighbors. Trained was done using loss weights of 1.0 for both energy and force labels.

\section{Active learning sampling parameters} \label{app:al}

\renewcommand{\thetable}{B\arabic{table}}
\setcounter{table}{0}  
To enhance configurational diversity during dataset generation, MD simulations were conducted in ASE~\cite{ASE} under perturbed thermodynamic conditions. Both temperature and density were varied dynamically throughout the simulations using a combination of linear ramping and sinusoidal modulation, following work done in Ref.~\cite{Smith2021}. This approach allowed the system to explore a broad range of equilibrium and non-equilibrium states relevant to uranium nitride.

\noindent At each MD time step \( t \), the temperature \( T(t) \) was updated as follows:
\[
T(t) = T_{\text{start}} + \left( \frac{t}{t_{\text{max}}} \right)(T_{\text{end}} - T_{\text{start}}) + T_{\text{amp}} \cdot \sin^2\left( \frac{t}{t_{\text{per}}} \right)
\]
The number density \( N/V(t) \) was modulated in an analogous manner:
\[
\frac{N}{V(t)} = \left( \frac{N}{V} \right)_{\text{start}} + \left( \frac{t}{t_{\text{max}}} \right)\left[\left( \frac{N}{V} \right)_{\text{end}} - \left( \frac{N}{V} \right)_{\text{start}} \right] + \left( \frac{N}{V} \right)_{\text{amp}} \cdot \sin^2\left( \frac{t}{t_{\text{per}}} \right)
\]
\noindent Density changes were implemented by scaling the simulation cell volume and atom positions by the ratio between the target and previous volume.

All MD simulations were carried out using the ASE Langevin thermostat with a timestep of 1.0~fs. Configuration snapshots were extracted every 10 steps. Only configurations with an ensemble uncertainty higher than the energy or force thresholds are stored in the dataset. We employed a threshold for the uncertainty in total energy of 28 eV/atom, for the mean uncertainty in force of 0.088 eV/\AA~, and for the maximum uncertainty in force of 0.024 eV/\AA.~The specific parameters used are summarized in Table~\ref{tab:sampling_params}.

\begin{table}[h!]
\caption{Molecular dynamics sampling parameters used for active learning.}
\label{tab:sampling_params}
\begin{ruledtabular}
\begin{tabular}{llp{8.5cm}}
\textbf{Parameter} & \textbf{Value/Range} & \textbf{Description} \\
\hline
\( T_{\text{start}} \) & 1000--5000 K & Initial temperature for thermal perturbation \\
\( T_{\text{end}} \) & 100--5000 K & Final temperature for ramped or oscillating MD \\
\( T_{\text{amp}} \) & 0--2000 K & Amplitude of temperature oscillation \\
\( t_{\text{per}} \) & 2--50 fs & Period of temperature/density oscillation \\
\( \left(\frac{N}{V}\right)_{\text{end}} \) & 11.78--16.08 g/cm\textsuperscript{3} & Final number density (controls volume) \\
\( \left(\frac{N}{V}\right)_{\text{amp}} \) & 0--0.1 g/cm\textsuperscript{3} & Amplitude of density modulation \\
\( t_{\text{max}} \) & 1000 ps & Total time of MD simulation \\
\end{tabular}
\end{ruledtabular}
\end{table}

\section{Validation of PBE functional: phonon bands} \label{app:dft}

Previous DFT calculation have shown that PBE+\textit{U} can reproduce the UN antiferromagnetic ordering seen in experiments at low temperatures. However, when using the Hubbard parameter, dynamical instabilities can be introduced in UN, manifesting itself as imaginary frequencies in the phonon spectra~\cite{Kocevski2022a}. 
To validate our functional, we calculated the phonon dispersions for a ferromagnetic uranium nitride on a $2 \times 2 \times 2$ supercell using PBE and PBE+\textit{U} with \textit{U}=1.85 eV functionals. The results are shown in Fig.~\ref{appa}, where one can observe the appearance of imaginary phonons with the PBE+\textit{U} functional.
 and verify that the PBE functional agrees the closest to the experimental data.

\renewcommand\thefigure{\thesection\arabic{figure}}
\setcounter{figure}{0}
\begin{figure*}[ht!]
\includegraphics[width=1\linewidth]{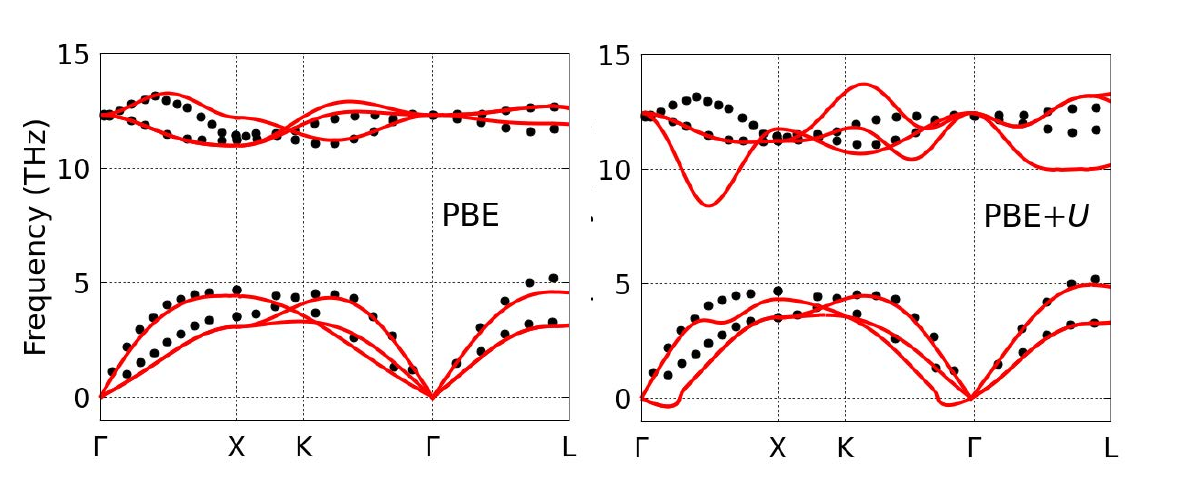}%
\caption{\label{appa} Phonon dispersion curves for UN. DFT data calculated (red lines) in a $2 \times 2 \times 2$ supercell using PBE and PBE+\textit{U} (\textit{U}=1.85 eV). Experimental data shown for comparison (black points).}
\end{figure*}

\newpage
\bibliography{reference}

\end{document}